\documentclass[10pt,conference]{IEEEtran}
\IEEEoverridecommandlockouts
\usepackage{cite}
\usepackage{amsmath,amssymb,amsfonts}
\usepackage[linesnumbered,ruled]{algorithm2e}
\usepackage{algpseudocode}
\usepackage{graphicx}
\usepackage{textcomp}
\usepackage{xcolor}
\usepackage{amsmath,amssymb,amsfonts}
\usepackage{float}
\usepackage{hyperref}
\usepackage{booktabs}
\usepackage{lipsum}
\usepackage{enumitem}
\usepackage{caption,graphicx}
\usepackage{subcaption}
\usepackage{cleveref}
\usepackage{amsmath} 
\usepackage{todonotes}
\usepackage{tabularx}
\usepackage{tabulary}
\usepackage{mdwtab}
\usepackage{tabularray}
\usepackage{multirow}
\usepackage[T1]{fontenc}
\usepackage{lmodern}
\usepackage{mdframed}
\mdfdefinestyle{graybox}{
    backgroundcolor=gray!20,
    linecolor=black,
    linewidth=1pt,
    roundcorner=5pt,
    skipabove=10pt,
    skipbelow=10pt,
    innertopmargin=6pt,
    innerbottommargin=6pt,
    innerleftmargin=6pt,
    innerrightmargin=6pt
}
\usepackage[most]{tcolorbox}
\usepackage[thinlines]{easytable}
\def\BibTeX{{\rm B\kern-.05em{\sc i\kern-.025em b}\kern-.08em
    T\kern-.1667em\lower.7ex\hbox{E}\kern-.125emX}}

\begin{document}

\title{Improving Image Data Leakage Detection in Automotive Software
%An Empirical Method for Data Leakage Detection in Automotive Perception Context
%An empirical study on data leakage detection in the automotive perception context
}
% \author{\IEEEauthorblockN{1\textsuperscript{st} Md Abu Ahammed Babu}
% \IEEEauthorblockA{\textit{Research \& Development} \\
% \textit{Volvo Car Corporation}\\
% Gothenburg, Sweden \\
% md.abu.ahammed.babu@volvocars.com}
% \and
% \IEEEauthorblockN{2\textsuperscript{nd} Sushant Kumar Pandey}
% \IEEEauthorblockA{\textit{Computer Science and Artificial Intelligence} \\
% \textit{University of Groningen}\\
% Groningen, Netherlands \\
% s.k.pandey@rug.nl}
% \and
% \IEEEauthorblockN{3\textsuperscript{rd} Darko Durisic}
% \IEEEauthorblockA{\textit{Research \& Development} \\
% \textit{Volvo Car Corporation}\\
% Gothenburg, Sweden \\
% darko.durisic@volvocars.com}
% \and
% \IEEEauthorblockN{4\textsuperscript{th} Ashok Chaitanya Koppisetty}
% \IEEEauthorblockA{\textit{Research \& Development} \\
% \textit{Volvo Car Corporation}\\
% Gothenburg, Sweden \\
% ashok.chaitanya.koppisetty@volvocars.com}
% \and
% \IEEEauthorblockN{5\textsuperscript{th} Miroslaw Staron}
% \IEEEauthorblockA{\textit{Department of Computer Science and Engineering} \\
% \textit{University of Gothenburg and Chalmers University of Technology}\\
% Gothenburg, Sweden \\
% miroslaw.staron@gu.se}
% }
\author
{\IEEEauthorblockN{Md Abu Ahammed Babu\textsuperscript{$1,3$}\qquad Sushant Kumar Pandey\textsuperscript{$2,3$}\qquad Darko Durisic\textsuperscript{$1$} 
\qquad   \\Ashok Chaitanya Koppisetty\textsuperscript{$1$} \qquad  Miroslaw Staron\textsuperscript{$3$}}
\IEEEauthorblockA{
\textsuperscript{$1$}Research \& Development,
 Volvo Car Corporation, Gothenburg, Sweden \\
 \textsuperscript{$2$}Bernoulli Institute for Mathematics, Computer Science and Artificial Intelligence, \\University of Groningen, The Netherlands \\
 \textsuperscript{$3$}University of Gothenburg and Chalmers University of Technology, Gothenburg, Sweden \\
}
\textbf{Email:} md.abu.ahammed.babu@volvocars.com, s.k.pandey@rug.nl, darko.durisic@volvocars.com, \\ashok.chaitanya.koppisetty@volvocars.com, miroslaw.staron@gu.se
}

% \author{
% \IEEEauthorblockN{Md Abu Ahammed Babu\textsuperscript{1,3}, Sushant Kumar Pandey\textsuperscript{2,3}, Darko Durisic\textsuperscript{1}, Ashok Chaitanya Koppisetty\textsuperscript{1}, Miroslaw Staron\textsuperscript{3}}
% \textsuperscript{1}Research \& Development, Volvo Car Corporation, Gothenburg, Sweden \\
% \textsuperscript{2}Bernoulli Institute for Mathematics, Computer Science and Artificial Intelligence, University of Groningen, The Netherlands \\
% \textsuperscript{3}University of Gothenburg and Chalmers University of Technology, Gothenburg, Sweden \\
% Email: \texttt{md.abu.ahammed.babu@volvocars.com, s.k.pandey@rug.nl, darko.durisic@volvocars.com, ashok.chaitanya.koppisetty@volvocars.com, miroslaw.staron@gu.se}

% }

\maketitle

%\textbf{Email:} md.abu.ahammed.babu@volvocars.com, s.k.pandey@rug.nl, darko.durisic@volvocars.com, ashok.chaitanya.koppisetty@volvocars.com, miroslaw.staron@gu.se

%\maketitle

%\renewcommand{\shortauthors}{Babu, et al.}

\begin{abstract}
Data leakage is a very common problem that is often overlooked during splitting data into train and test sets before training any ML/DL model. The model performance gets artificially inflated with the presence of data leakage during the evaluation phase which often leads the model to erroneous prediction on real-time deployment. However, detecting the presence of such leakage is challenging, particularly in the object detection context of perception systems where the model needs to be supplied with image data for training. In this study, we conduct a computational experiment on the Cirrus dataset from our industrial partner Volvo Cars to develop a method for detecting data leakage. We then evaluate the method on another public dataset, “Kitti”, which is a popular and widely accepted benchmark dataset in the automotive domain. The results show that thanks to our proposed method we are able to detect data leakage in the “Kitti” dataset, which was previously unknown.
\end{abstract}

\begin{IEEEkeywords}
Data leakage detection, object detection, YOLOv7, Automotive perception systems
\end{IEEEkeywords}

\section{Introduction}
\label{sec: introduction}
Autonomous driving (AD) is an automotive software system constructed by combining multiple perception sub-systems \cite{kiran2021deep}. The autonomous perception sub-systems detect (perceive) objects by using data collected from the operational design domain (ODD) using different types of sensors e,g. LiDAR, Radar, Camera, and Ultrasound sensors. One of the sources of data is the camera, which is used in object detection (OD) scenarios in autonomous driving research, as it plays a crucial role in determining the safety of self-driving vehicles \cite{gupta2021deep}. This includes quick and accurate identification of potential hazards in the surrounding traffic, as well as detecting traffic signs and road conditions for effective route planning. Overall, object detection is a long-term research priority in the development of autonomous driving technology \cite{rashed2021generalized}.

In recent years, the development of automotive perception systems has revolutionized the automotive industry, paving the way for advanced driver-assistance systems and autonomous vehicles. These systems rely heavily on image data for tasks such as vehicle detection, vehicle model recognition, and component recognition \cite{sun2006road}. However, the issue of data leakage during the splitting of image data can pose a significant threat to the performance and reliability of these crucial tasks \cite{ma20233d}. %In this paper, we will explore the impact of data leakage in the context of automotive perception systems, highlighting the potential challenges and implications for the development and deployment of such systems. We will also delve into potential solutions and best practices to mitigate the effects of data leakage, ensuring the robustness and accuracy of automotive perception systems.

%Modern object detection techniques have been evolving fast with the significant development of deep neural networks (DNN). The deep learning (DL) based object detectors \cite{redmon2016you} mostly use supervised or semi-supervised learning where the model learns from train image data and the performance of the model is later evaluated through various performance metrics (e,g, mAP, AUC, etc.) with test image sets. The goal of the training process is to make the model able to detect object(s) in images with the help of previous knowledge. One of the main challenges to designing an accurate and robust object detection system is the potential of data leakage \cite{silva2022machine}.

In the automotive domain, the V-model is commonly used for software development \cite{rana2015software}. During the development of automotive software, it is important to follow the classical machine learning workflow. According to Pandey et al., \cite{pandey2023data}, the training phase of the machine learning workflow overlaps with the implementation and unit testing phases of the V-model, while performance evaluation aligns with integration and integration testing. However, the authors mentioned during training a model potential for overestimating the performance of a model, as test data information may already be present in the training data, leading to data leakage issues. For our industrial partner, the safety of the automotive perception system is paramount and therefore requires state-of-the-art methods to detect such problems in the datasets used to train their model.

In general, data leakage occurs when a subset of training data is used as well (leaked) in the testing dataset \cite{baby2017literature}.
This can inflate the model performance in the testing scenario, as the model has been trained and tested on this subset. This inflated performance is not observed in real-life applications, which means that the system can perform significantly worse \cite{kernbach2022foundations}, e.g., leading to risks in real traffic situations. It is particularly important for vision perception systems, where images and video feed systems can include similar (although not identical) images. Therefore, two (or more) consecutive images and frames can differ very little, so the random split of the data can contain images that are similar but not identical. This can lead to overly positive performance results of the trained classifiers \cite{cawley2010over}. Hence, splitting the data becomes a crucial step in training and evaluating models for autonomous driving \cite{li2017secure}.

Unfortunately, detection of whether data leakage occurred during splitting is hard for the image data \cite{drobnjakovic2022abstract, ji2023critical}, due to factors like image similarity, context similarity \cite{apicella2024don}, and semantic similarity \cite{andre2012learning}. In this study, we propose a method for data leakage detection, particularly in the context of OD tasks of automotive perception systems. The study addresses the following research questions (RQs):

%\textbf{RQ1: }How effective is the proposed method in detecting data leaks in object detection systems?

%\textbf{RQ1.1:} How effective is the proposed method in detecting data leaks in a new automotive dataset?

\textbf{RQ1:} How does incremental data leakage impact the object detection performance?

%\textbf{RQ2:} Which performance measure helps to indicate the presence of data leakage?

\textbf{RQ2:} How to detect the presence of data leakage in the existing split?

\textbf{RQ3:} How effective is the proposed method in detecting data leakage in automotive datasets?

%\textbf{RQ3:} How does data leakage affect the resilience of the model?
%\textbf{RQ3:} To what extent does incremental data leakage affect the model's resilience for autonomous driving?

Answering these research questions is crucial for the automotive Original Equipment Manufacturers (OEM) that work intensively on developing autonomous driving technologies, like our industrial partner. In order to assure the safety of their vehicles, Volvo's software engineers need to understand how reliable and secure their image perception systems are in real-world scenarios. Autonomous vehicles rely heavily on these systems to detect and interpret their surroundings accurately, making the detection of data leakage or erroneous predictions in machine-learning models a critical aspect of ensuring vehicle safety. For Volvo, ensuring the robustness of these perception systems is essential to maintaining the high safety standards that define the company's reputation, and any issues related to image recognition or data integrity could directly impact the safety of their autonomous driving solutions.

The next sections of the paper are organized in a way where Section \ref{sec: background} explores existing literature related to data leakage, its definition, and its consequences. Section \ref{sec: research design} explains the methodology of this empirical study. Section \ref{sec: results} shows the findings of this study and a method of data leakage detection will be proposed in Section \ref{sec: proposed method} based on the findings. Finally, Section \ref{sec: evaluation} presents the evaluation of the proposed method for data leakage detection on a popular dataset. Section \ref{sec: discussion} contains a thorough discussion of the findings and evaluation, followed by Section \ref{sec: conclusion}, which provides a conclusion of the study and also points to the possibility of future research scope.

\section{Background and related work}
\label{sec: background}
\emph{Data leakage} is a situation during the training process where a feature that is later found to be associated with the outcome is used as a predictor \cite{silva2022machine}. It occurs when information about the outcome is inadvertently included in the data used to build the model \cite{silva2022machine}. For example, when the same data point is used for both training and testing of the machine learning model. A few other reasons for data leakage could be related to the pre-processing of data such as imputing average to fill up the missing values, de-seasonalization which utilizes monthly averages of time-series data, or using mutually dependent variables to predict one using the other \cite{hussein2022rainfall}. Data leakage \cite{baby2017literature} is a common but crucial problem that is often overlooked during the development and deployment of supervised or semi-supervised ML/DL models. The majority of the training process relies on object identity when creating the splits -- it is enough that exactly the same image is not included in both sets. However, the presence of this problem could be even more hazardous in safety-critical systems like autonomous driving where images are not identical but could be extremely similar. For example, when two consecutive frames from a driving video feed are included in train and test sets respectively. These frames are not identical, but very similar. Although the data leakage problem is known to the ML/DL research community, the ways of identifying the presence or how to avoid this issue need more attention.

Many experts believe that data leakage is a major issue in machine learning that also contributes to the problem of irreproducibility \cite{sculley2015hidden}. One can argue that the definition of data leakage should be broadened to encompass any type of information flow between data used at different stages of the machine learning pipeline, such as the availability of validation set information during training \cite{gotz2022critical}. Although this kind of leakage may not necessarily improve performance on an independent test set, it is still a problem similar in nature to classical data leakage. As a result, detecting data leakage can be challenging, particularly when there are multiple processing steps or statistical information extracted during pre-processing \cite{gotz2022critical}.

\begin{figure}[!h]
    \centering
    \begin{subfigure}[b]{0.45\textwidth}
	   \centering
	   \includegraphics[width=\linewidth]{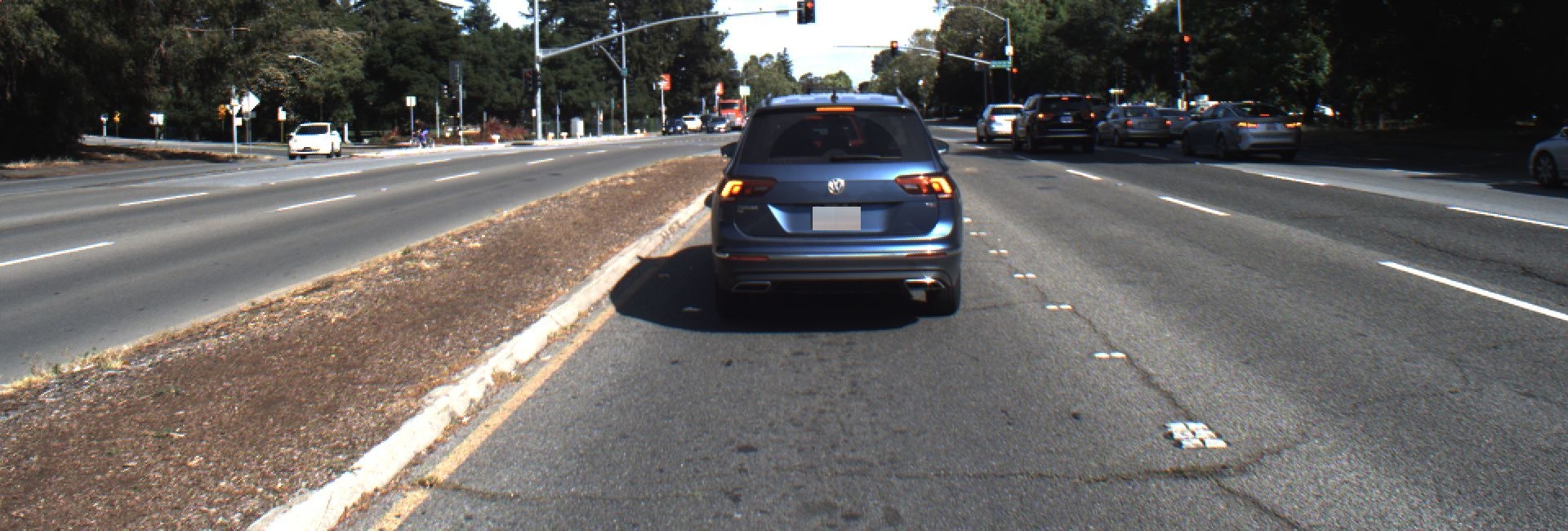}
	   \caption{\scriptsize Train image}
	   \label{image:train_image}
	\end{subfigure}
    \hfill
    \begin{subfigure}[b]{0.45\textwidth}
	   \centering
	   \includegraphics[width=\linewidth]{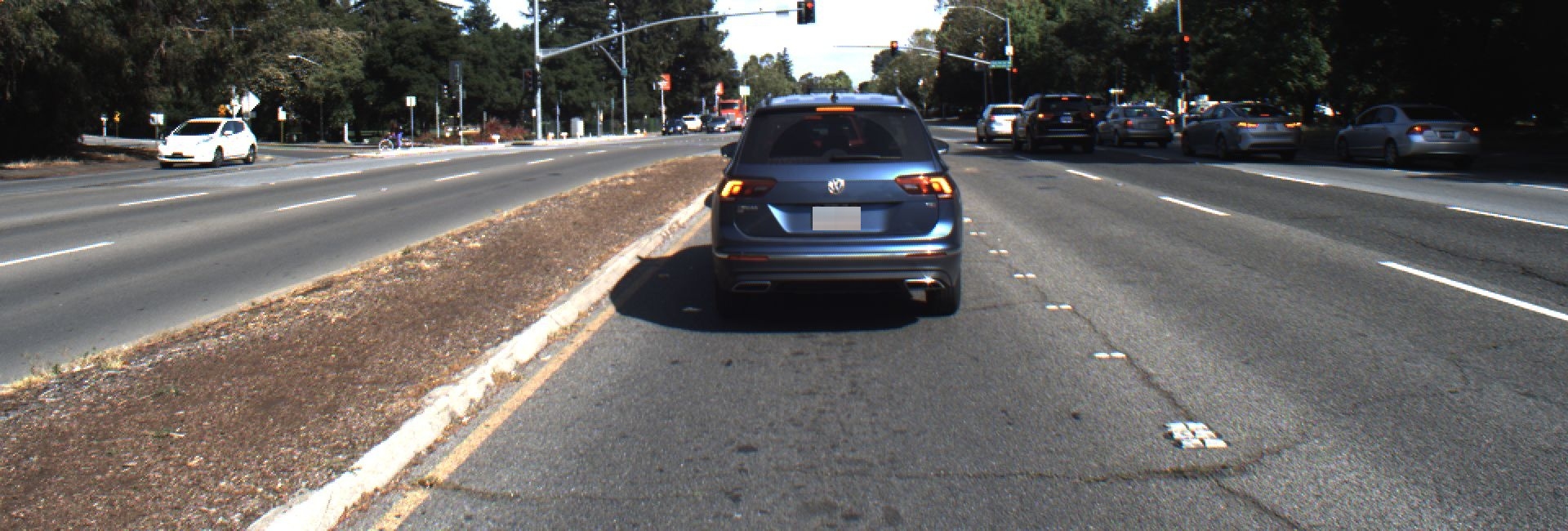}
	   \caption{\scriptsize Test image}
	   \label{image:test_image}
	\end{subfigure}
    \caption{An example of target leakage through similar images present in both train and test datasets}
    \label{fig: An example of target leakage through similar images present in both train and test}
\end{figure}

In practice, data could be leaked through any common feature(s) (also called target leakage) even if developers take measures to ensure no data occurs repeatedly in both train and test sets \cite{kernbach2022foundations}. Target leakage occurs when for example, an image like \ref{image:train_image} belongs to train data and a very similar image (but not exactly the same) to it like \ref{image:test_image} is present in the test data. Then the model will learn the correlation between the present objects and the remaining background pattern of the image instead of learning the unique properties of objects. Thus, the detection performance on the test data would be erroneously influenced and hence, the actual performance of the model will not be demonstrated in the testing phase. To avoid data leakage in deep learning model training, data splitting should be done carefully \cite{rouzrokh2022mitigating}. Different splitting techniques need to be examined and evaluated to find which one is the most appropriate and most likely to guarantee the absence of leakage or data exposure from the train set to the test set or vice versa. According to a study on the effects of alternative splitting rules on image processing, the classification accuracy does not differ significantly for varying splitting rules/techniques \cite{zambon2006effect}.

%\todo[inline]{You need to explain what "target leakage" is. I also recommend to provide a motivating example where you actually show two consecutive frames with similar images.}

%For a model to perform effectively with new data in real-world scenarios, it is important to prevent data leakage during its development. Data leakage can hinder the model's ability to make accurate predictions when faced with new data. Therefore, it is crucial to prevent data leakage while training the model \cite{tusar2022detecting}. Data leakage could occur either for incorrect data split or leakage of direct target data label \cite{yagis2021effect}. Data split is one of the major causes of leaking data to both validation and test sets \cite{wen2020convolutional} and hence needs to be done carefully. %Hence it can be avoided by creating a proper split, late split, and applying data pre-processing steps such as cross-validation before splitting the data into the train-test set.
%\todo[inline]{The paragraph above does not fit into the related work section.}
Many studies have shown that despite having a CNN model with high performance reported, making the model generalizable is more challenging due to possible data leakage introduced during cross-validation of the model. The study conducted by Yagis et al. \cite{yagis2021effect} reports that the performance of the deep learning model might be overly optimistic due to potential data leaks caused by either improper or late data split. The authors explored previous studies in the medical field related to classifying MRI images and found that the test accuracy gets erroneously inflated by 40-55\% on smaller datasets and 20-45\% on larger datasets due to incorrect slice-level cross-validation, which causes data leakage. Another study on assessing the impact of data leakage on the performance of machine learning models in the biomedical field \cite{bussola2021ai} showed that the predictive scores can be inflated up to 41\% even with a properly designed data analysis plan (DAP). The authors replicated the experiments for 4 classification tasks on 3 histopathological datasets. Another study on the application of deep learning in optical coherent tomography (OCT) data has found that the classification performance may inflate by 0.07 up to 0.43 for models tested on datasets with improper splitting.

The existing literature clearly highlights “data leak” as a crucial problem and one of the major impediments in the way of having a generalizable machine learning/deep learning model. Some studies also concluded that the occurrence of data leakage often creates the irreproducibility issues \cite{kapoor2023leakage, wen2020convolutional} of the previous research, and some may cause incorrectly highly inflated results \cite{shim2021inflated, pulini2019classification}. The authors of \cite{apicella2024don} categorized different types of data leakage in ML based on the possible reasons behind its occurrence. In addition, they also emphasized on the importance of addressing data leakage for robust and reliable ML applications. Yang et. al. have developed a static analysis approach \cite{yang2022data} to detect common forms of data leakage in data science code by analyzing 1000 public notebooks. The approach yields 97.6\% precision and 67.8\% recall with an overall accuracy of 92.9\% in detecting preprocessing leakage (detects 262 out of 282 potential leakages). Unfortunately, this method of detecting data leakage is limited to what can be seen in the static code, typically in a data science notebook. It may not be effective in more complex or adversarial settings where different coding practices are used \cite{yang2022data}. The detection of leakage in image recognition contexts, such as for OD tasks in AD which is very crucial for passenger safety, appears to be under-explored in the existing literature. Since the appearance of image data is different compared to numerical data in terms of many properties like visualization, luminance, background information etc., the existing data leakage methods for numeric/code data cannot serve the purpose either. Moreover, the Clever Hans effect \footnote{“Clever Hans effect” is used in psychology to describe when an animal or a person senses what someone wants them to do, even though they are not deliberately being given signals \cite{de2016we}.} \cite{lapuschkin2019unmasking} might be another constraint in leakage detection in cases where image data is used. Clever Hans happens when the trained ML model actually exploits features and correlation patterns with the target class and may mislead the model to distinguish between the classes based on the surrounding features such as light condition, background, etc. \cite{apicella2024don}. Hence, finding a method to detect data leakage in such cases appears to be an inescapable task. Data leak detection and prevention is also essential to ensure the safe and reliable operation of safety-critical applications like autonomous driving.

In summary, the existing literature identifies “data leakage” as a very common but crucial problem in training ML/DL models but very few have found a way of detecting potential leakage of data in some particular context. However, techniques for data leak detection in the field of image recognition systems and operations like OD are yet to be explored.

%In the context of automotive perception systems, the splitting of image data plays a crucial role in tasks such as vehicle detection, model recognition, and component recognition \cite{liang2018car}. Although extensive research has been done on defining data leaks and quantifying the impact of data leakage for various time-series data and predictive models, there is a scarcity of similar works in the field of image processing and computer vision. There was no study found related to data leak and their impact, particularly in the field of autonomous driving up to this time and the author's utmost knowledge. Hence, the necessity of conducting studies in this field to focus on data leakage, its impact on autonomous driving, and how to detect and overcome/avoid it felt like a significant move.

\section{Research design}
\label{sec: research design}
Our study has been done in the form of a computational experiment in a controlled setting. That means all the experiment steps were executed in a fixed hardware configuration and carefully monitored to avoid any spurious mix of data and/or results. A server equipped with an Intel Core-i7 CPU running at 3.70 GHz and 32.0 GB of RAM with an extra NVIDIA GeForce RTX 4090 GPU was used for the experiment. The replication package with necessary scripts and instructions is made available\footnote{\url{https://figshare.com/s/acb5023a7fc3b99b9051}}.
\begin{figure*}
    \centering
    \includegraphics[width=0.8\linewidth]{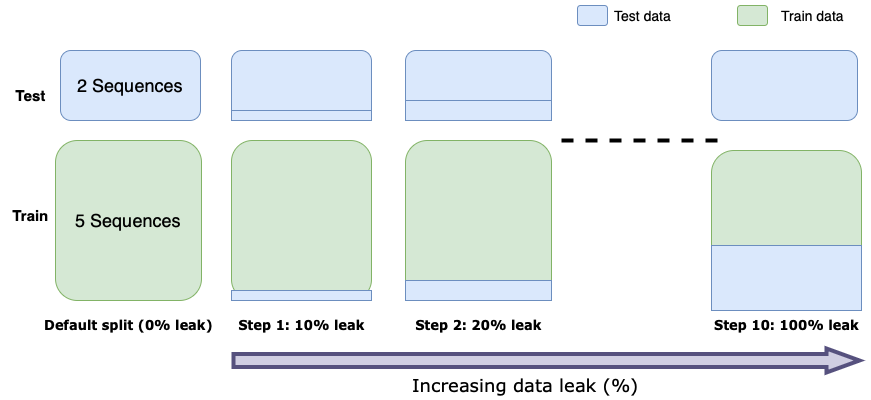}
    \caption{Illustration of the data leakage steps.}
    \label{Fig: Illustration of the data leakage steps}
\end{figure*}
The Volvo Cars Cirrus dataset was used in this experiment because it was collected in a real-world situation with a natural setup. The Cirrus dataset \cite{wang2021cirrus} is an open-source dataset for autonomous driving with sequences of images collected in the Silicon Valley area. Cirrus contains 6,285 RGB images in 7 separated sequences/folders from both high-speed highway driving and low-speed urban road scenarios. Each of the sequences represents different driving scenarios as well as different geographical locations which makes the dataset unique from the other such datasets in the automotive domain. This uniqueness also makes this dataset perfect for using in this study. %Unfortunately, the 3D annotations that came with the dataset could not be used due to improper calibration matrices and insufficient supporting information on how to translate the annotations. 
We have used the corresponding 2D annotations \footnote{The 2D annotations are available at \url{https://developer.volvocars.com/resources/cirrus}} which was also used in a previous study \cite{babu2024impact}.

%YOLOv7 model\todo{what do you mean by this statement -- it is not a sentence nor a heading, it disrupts the flow of text.}. 
The ultimate task focused during this study is the object detection (OD) of autonomous vehicles (AV). The YOLOv7 \cite{wang2022yolov7}, which is one of the latest editions in the YOLO (You Only Look Once) family, was used for this task. YOLO models are renowned for their high speed of operation with consistent accuracy \cite{li2020yolo}, which is the main reason behind choosing YOLOv7 for this study. They are also frequently used in embedded software systems for these reasons. 

%\subsection{Experiment}
In the initial step, the dataset was split into train and test sets to prevent data leakage. The first five(5) sequences were chosen as the training set, and the remaining two(2) were used for testing. This leads to roughly a 70-30 train-test split ratio, which is often considered standard practice. This setting of the “train” and “test” data was kept the same during the whole experiment. Moreover, the chosen sequences also come from different driving scenarios and thus ensures inclusion of all scenario representative images in the train set. This initial split has been done cautiously so that no data leakage can happen throughout the experiment.

Once the train and test datasets are fixed, the next steps are intentionally leaking data from the test to the train dataset in an incremental manner. In every step, we chose to leak 10\% of the test images to the training dataset and replaced the same number of images from the train to keep the train-test ratio consistent. A step size of 10\% is chosen as it provides a good visualization of the impact of leakage on the test performance after every step. %Choosing a smaller size like 5\% will unusually increase the number of steps and a 20\% step size may not unravel the actual circumstance.\todo{This sentence is a bit speculative, is there any backing for this discussion of 5 and 20\%?}
A graphical illustration of how the steps were performed is shown in Figure \ref{Fig: Illustration of the data leakage steps}.

%\subsection{Evaluation}
The total number of images was 1,790 in the test dataset. Hence, in the first step, 179 images (10\% of test data) were randomly chosen to be copied to the training dataset and replaced with the same number of images randomly. To avoid random bias, the whole process is repeated 10 times which creates 10 different versions of the training dataset. After every repetition, the YOLOv7 model was trained on the new train set and evaluated on the same test set again. In the end, the mean of the performance scores were reported. The number of images leaked in every step is tabularized in table \ref{tab: Number of duplicate images in both train and test dataset in each step} for ease of understanding.

\begin{table}[htbp]
\scriptsize
%\footnotesize
    \caption{Number of duplicate images in both train and test dataset in each step}
    \begin{center}
    %\resizebox{\columnwidth}{!}{%
        \begin{tabular}{|c|c|c|}
        \hline
        \textbf{Steps} & \textbf{Percentage of leakage} & \textbf{Number of duplicate images} \\
        \hline
        0 & 0\% & 0 \\
        %\hline
        1 & 10\% & 179 \\
        %\hline
        2 & 20\% & 358 \\
        %\hline
        3 & 30\% & 537 \\
        %\hline
        4 & 40\% & 716 \\
        %\hline
        5 & 50\% & 895 \\
        %\hline
        6 & 60\% & 1,074 \\
        %\hline
        7 & 70\% & 1,253 \\
        %\hline
        8 & 80\% & 1,432 \\
        %\hline
        9 & 90\% & 1,611 \\
        %\hline
        10 & 100\% & 1,790 \\
        \hline
        \end{tabular}
        %}
        \label{tab: Number of duplicate images in both train and test dataset in each step}
    \end{center}
\end{table}

For performance comparison, we take all four available performance metrics which come as default with the YOLOv7 model training into consideration every time. Among them, both mAP and F1-score are widely adopted, particularly for OD tasks, as both of them take precision and recall into account and combine them to generate a balanced score \cite{al2024ensemble, casas2023assessing}. Additionally, we computed perceptual hash (pHash) distances for every train-test image pair to assess the perceptual/visual similarities between the training and testing images. pHash is a renowned method to compare visual similarity between images. It is a technique used to generate hash values that represent the content of an image in a way that is resilient to minor transformations such as scaling, rotation, or color adjustments \cite{zauner2010implementation}. Unlike cryptographic hashes, which change drastically with even the smallest alteration to the input, pHash creates similar hash values for visually similar images, making it well-suited for comparing image content \cite{monga2006clustering}. The pHash distance of two images is basically the hamming distance of their perceptual hash values. Hence, the pHash distance of two images is 0 means the images are almost identical to each other. In this study, pHash was utilized to assess the similarity between the training and testing image datasets, allowing for an analysis of how visually consistent these datasets are. This method is particularly effective for detecting near-duplicate images or variations across the datasets, ensuring that the models are evaluated on truly distinct data.

\section{Results}
\label{sec: results}
Table \ref{tab: Average results summary after 10 iterations of each step} provides a performance summary of the averages of precision, recall, mAP, and F1-score for every data leak step.

\begin{table}[!ht]
	\centering
	\scriptsize
 
 %\scriptsize
	\caption{Average results summary after 10 iterations of each step}
	\begin{tabular}{| c|p{1.4cm}|c|c|c|c| }
		 \hline
        \textbf{Steps} & \textbf{Percentage of leakage} & \textbf{Precision} & \textbf{Recall} & \textbf{mAP} & \textbf{F1-score} \\
        \hline
        % \hline
        0 & 0\% & 0.553 & 0.469 & 0.486 & 0.49 \\
        %\hline
        1 & 10\% & 0.622 & 0.563 & 0.595 & 0.57 \\
        %\hline
        2 & 20\% & 0.690 & 0.628 & 0.701 & 0.64 \\
        %\hline
        3 & 30\% & 0.761 & 0.641 & 0.701 & 0.67 \\
        %\hline
        4 & 40\% & 0.764 & 0.669 & 0.736 & 0.70 \\
        %\hline
        5 & 50\% & 0.831 & 0.680 & 0.760 & 0.72 \\
        %\hline
        6 & 60\% & 0.786 & 0.722 & 0.783 & 0.74 \\
        %\hline
        7 & 70\% & 0.787 & 0.739 & 0.791 & 0.75 \\
        %\hline
        8 & 80\% & 0.820 & 0.757 & 0.815 & 0.77 \\
        %\hline
        9 & 90\% & 0.843 & 0.769 & 0.829 & 0.78 \\
        %\hline
        10 & 100\% & 0.831 & 0.800 & 0.835 & 0.79 \\
        \hline
        
\end{tabular}
	\label{tab: Average results summary after 10 iterations of each step}
\end{table}

\begin{figure}
    \centering
    \includegraphics[width=1\linewidth]{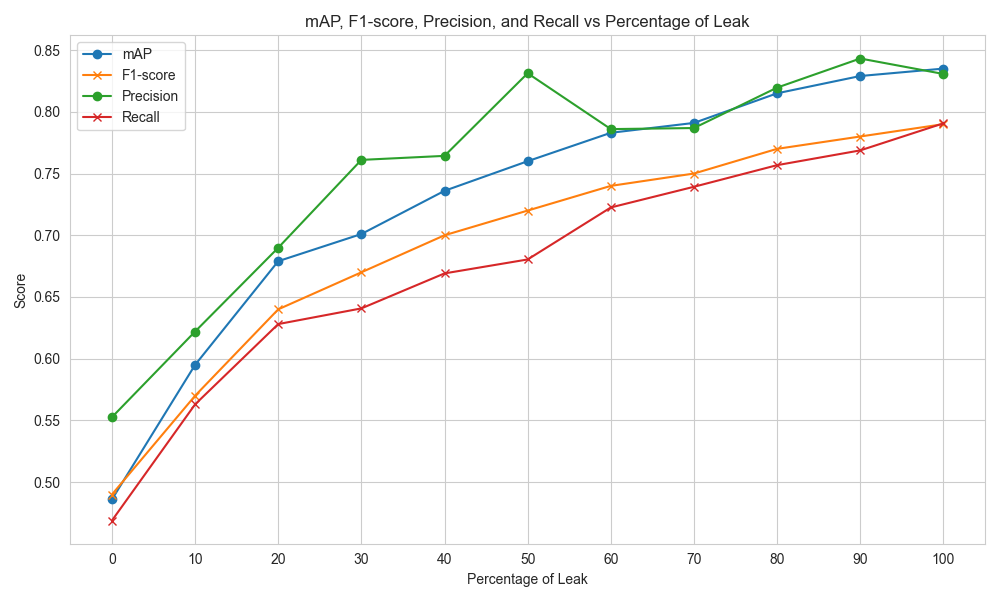}
    \caption{Results summary graph}
    \label{Fig: Results summary graph}
\end{figure}

From the table, the increase in all four performance metrics is clearly visible, as expected. However, how the scores increased is clearly visible in Figure \ref{Fig: Results summary graph} with precision, recall, mAP, and F1-score for every step 0 -- 100\% data leak. The graph shows that there was a sharper increase of all four metrics for 0 -- 20\% data leak. This answers our RQ1 about the impact of incremental data leaks on performance. For leakage of above 20\% data, the performance steadily kept increasing, but with a lower rate, especially in the case of mAP and F1-score. As both mAP and F1-score follow a regular increase pattern with the increase of leakage percentage, these two alone or together can be used as indicator(s) of data leakage in the existing data split. However, the values do not increase at the same rate after a 70-80\% data leakage. In other words, the increase rate gets lower with a higher percentage of data leakage.

The pHash distance values of all the train-test image pairs were also calculated for the Cirrus datasets and are reported in Table \ref{tab: pHash distance of the ‘Cirrus’ dataset}. Typically, a threshold of 10 or less is commonly used to determine if two images are perceptually similar, meaning that the images differ in minor details\cite{zauner2010implementation} thus the pHash distances of up to 10 were only reported in the table. The lowest pHash distance found is 4 which occurs only for two image pairs. The second lowest pHash distance found is 6 and it also occurs only for three image pairs. This finding ensures that the initial split of Cirrus datasets was leakage free and there were hardly similar images among the train and test datasets.

\section{Proposed method}
\label{sec: proposed method}
The performance summary graph in the previous section shows that performance does not increase at a regular rate, particularly in terms of mAP and F1 scores. This led us to creating the method based on this relative increase. To further examine this rate of change over each step of incremental data leakage percentage, the relative rate of performance increase was calculated according to Equation \ref{eq_relative_increase_rate}. Table \ref{tab: The relative increase rate of mAP and F1-score} shows how the values of mAP and F1-score relatively increased with the incremental data leak from 0 to 100\% of data leakage in the training dataset.
\begin{table}[htbp]
%\scriptsize
%\footnotesize
    \caption{pHash distances of train-test image pairs of ‘Cirrus’ dataset}
    \begin{center}
        \begin{tabular}{|c|c|}
        \hline
        \textbf{pHash distances} & \textbf{\# of Occurances} \\
        \hline
        % Train-test ratios & 0.9880 > 0.01 \\
        % \hline
        4 & 2 \\
        %\hline
        6 & 3 \\
        %\hline
        8 & 15 \\
        %\hline
        10 & 27 \\
        \hline
        Total & 47 \\
        %\hline
        \hline
        \end{tabular}
        \label{tab: pHash distance of the ‘Cirrus’ dataset}
    \end{center}
\end{table}

\begin{equation}
Relative\_increase = \frac{current\_value - previous\_value}{previous\_value}
\label{eq_relative_increase_rate}
\end{equation}

\begin{table}[htbp]
\scriptsize
%\footnotesize
    \caption{The relative increase rate of mAP and F1-score}
    \begin{center}
    %\resizebox{\columnwidth}{!}{%
        \begin{tabular}{|p{0.7cm}|p{1.5cm}|p{2.2cm}|p{2.4cm}|}
        \hline
        \textbf{Steps} & \textbf{Percentage of leakage} & \textbf{Relative increase (mAP)} & \textbf{Relative increase (F1-score)} \\
        \hline
        0 & 0\% & 0\% & 0\% \\
        %\hline
        1 & 10\% & 22.4\% & 16.3\% \\
        %\hline
        2 & 20\% & 14.1\% & 12.3\% \\
        %\hline
        3 & 30\% & 3.2\% & 4.7\% \\
        %\hline
        4 & 40\% & 5.0\% & 4.5\% \\
        %\hline
        5 & 50\% & 3.3\% & 2.9\% \\
        %\hline
        6 & 60\% & 3.0\% & 2.8\% \\
        %\hline
        7 & 70\% & 1.0\% & 1.4\% \\
        %\hline
        8 & 80\% & 3.0\% & 2.7\% \\
        %\hline
        9 & 90\% & 1.7\% & 1.3\% \\
        %\hline
        10 & 100\% & 0.7\% & 1.3\% \\
        \hline
        \end{tabular}
        %}
        \label{tab: The relative increase rate of mAP and F1-score}
    \end{center}
\end{table}

\begin{figure}
    \centering
    \includegraphics[width=1\linewidth]{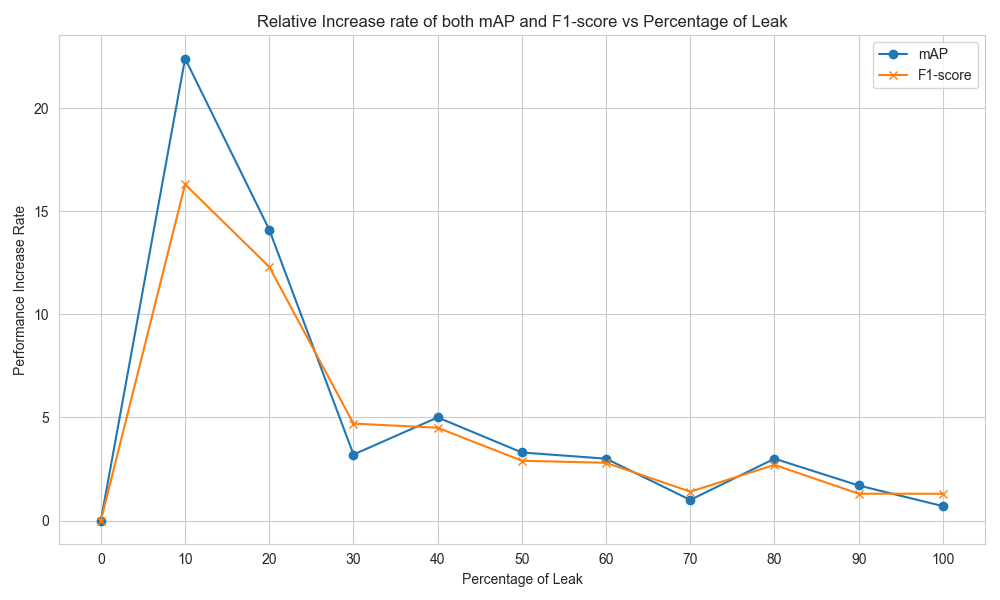}
    \caption{The relative performance increase rate}
    \label{Fig: The relative performance increase rate}
\end{figure}

The values indicate that both the mAP and F1-score swiftly increased with 22.4\% and 14.1\% for mAP, 16.3\%, and 12.3\% for F1-score in the first couple of steps (with 10\% and 20\% data have been leaked). This rate of increase was consistent over the next 5-6 steps for both the performance metrics (in between 2-5\% relative increase). However, in the last two steps, where the percentage of leakage was high (more than 80\%), the relative increase rate was below 2\% for both mAP and F1-score. This shows that the performance does not increase highly when a high percentage of data leaks occur during splitting. This answers our RQ3 of how to detect the presence of data leakage in the existing split. A graph has also been shown in Figure \ref{Fig: The relative performance increase rate} to expose the variation of the relative rate of performance increase.

Based on the unusual pattern of relative performance increase rate, we propose a method for detecting the presence of data leakage in a current data split. The proposed method tells that intentional leakage of data in a systematic manner like ours can indicate whether an arbitrary data split suffers from data leakage or not. The performance scores of the model trained with the incrementally leaked dataset are used to calculate the relative increase rate. The presence of data leakage occurs if the relative increase of performance is low ($\le$5\%) during the first two steps (i.e., when leaking 10\% and 20\% respectively). On the contrary, if the relative performance rate sharply increases with the increased data leakage, no initial data leakage is present in the existing data split. The method is explained through an algorithmic notation in Algorithm \ref{algo1}. This method can be utilized by practitioners whenever there is doubt about the presence of potential data leakage in their existing split since data leakage detection is not a straightforward task in the examined domain of OD.

%\todo[inline]{The proposed method is good, but it could be a bit more explicit: do this, measure that, if X then Y; in other words more actionable. }

% \hspace*{\algorithmicindent} \textbf{Input} \\
% \hspace*{\algorithmicindent} \textbf{Output}
% \begin{algorithm}[!httb]
% 	\caption{Data leakage detection in an arbitrary split of image data}
% 	\label{algo1}
% 	\textbf{Inputs} $\gets$ Train images (Tr$_{d}$), Test images (Te$_{d}$) \\
%     \textbf{Output} $\rightarrow$ Data leakage detected (Yes / No)
% 	\\
%  \hrulefill

%     $leakage\_percentage \gets 10\%$\;
    
%     $data\_leakage\_presence \gets FALSE$;
    
%     \While {$leakage\_percentage \leq 20\%$} {Calculate $relative\_increase\_rate, R_i$; 
%    \\ \eIf{$R_i \leq 5\%$}{$data\_leakage\_presence \gets TRUE$\; \\
%    /* Presence of potential data leakage detected */
%    }{
%         $Continue$;\;
%     }
%     }
%     %\ENDWHILE
% \end{algorithm}

\begin{algorithm}[!httb]
	\caption{Data leakage detection in an arbitrary split of image data}
	\label{algo1}
	\textbf{Inputs} $\gets$ Train images (Tr$_{d}$), Test images (Te$_{d}$) \\
    \textbf{Output} $\rightarrow$ Data leakage detected (Yes / No)
	\\
 \hrulefill

    $leakage\_percentage \gets 10\%$\;
    
    $data\_leakage\_presence \gets FALSE$\;
    
    \While {$leakage\_percentage \leq 20\%$} {
        Calculate $relative\_increase\_rate, R_i$;\;
        \eIf{$R_i \leq 5\%$}{
            $data\_leakage\_presence \gets TRUE$;\; 
            /* Presence of potential data leakage detected */
        }{
            $Continue$;\;
        }
    }
\end{algorithm}

\section{Evaluation}
\label{sec: evaluation}
\begin{figure}
    \centering
    \includegraphics[width=1\linewidth]{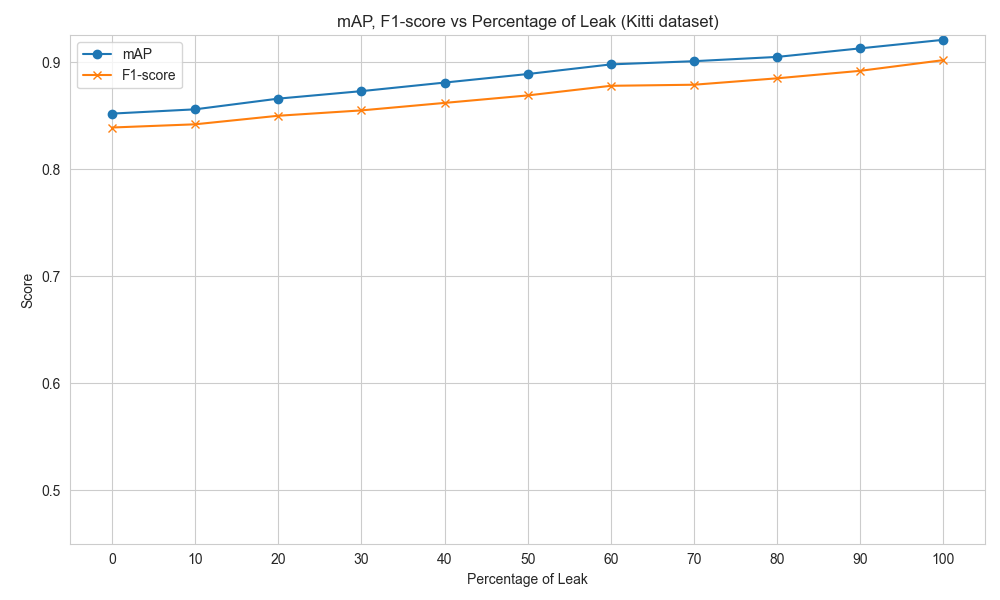}
    \caption{Evaluation results summary on kitti}
    \label{Fig: Evaluation results summary on kitti}
\end{figure}

To evaluate the proposed method, we have replicated the experiment on the Kitti dataset \cite{geiger2013vision}, which is one of the most popular benchmark datasets in the AV research field and widely adopted for testing and benchmarking new OD models in the automotive field. Kitti has two separate image folders called “train” and “test”. The train folder contains 7,481 images along with annotations of 9 object classes, and the test images do not have their corresponding annotation files available with them. We have chosen to split the original “train” data of Kitti to get “train” and “test” data from that with a 70-30 ratio. That leads to having the first 5,231 images in the “train” and the remaining 2,250 images in the “test” dataset. The evaluation results are summarized in Table \ref{tab: Evaluation results of the proposed method in kitti dataset} and also drawn in the line graph in Figure \ref{Fig: Evaluation results summary on kitti}.

\begin{table*}[htbp]
\scriptsize
%\footnotesize
    \caption{Evaluation results of the proposed method in kitti dataset}
    \begin{center}
    \resizebox{\linewidth}{!}{%
        \begin{tabular}{|c|c|c|c|c|c|}
        \hline
        \textbf{Steps} & \textbf{\% of leakage} & \textbf{mAP} & \textbf{F1-score} & \textbf{Relative increase (mAP)} & \textbf{Relative increase (F1-score)} \\
        \hline
        % Train-test ratios & 0.9880 > 0.01 \\
        % \hline
        0 & 0\% & 0.852 & 0.839 & 0 & 0 \\
        %\hline
        1 & 10\% & 0.856 & 0.842 & 0.47\% & 0.36\% \\
        %\hline
        2 & 20\% & 0.866 & 0.850 & 1.17\% & 0.95\% \\
        %\hline
        3 & 30\% & 0.873 & 0.855 & 0.81\% & 0.59\% \\
        %\hline
        4 & 40\% & 0.881 & 0.862 & 0.92\% & 0.82\% \\
        %\hline
        5 & 50\% & 0.889 & 0.869 & 0.91\% & 0.81\% \\
        %\hline
        6 & 60\% & 0.898 & 0.878 & 1.01\% & 1.04\% \\
        %\hline
        7 & 70\% & 0.901 & 0.879 & 0.33\% & 0.11\% \\
        %\hline
        8 & 80\% & 0.905 & 0.885 & 0.44\% & 0.68\% \\
        %\hline
        9 & 90\% & 0.913 & 0.892 & 0.88\% & 0.79\% \\
        %\hline
        10 & 100\% & 0.921 & 0.902 & 0.88\% & 1.12\% \\
        \hline
        \end{tabular}
        }
        \label{tab: Evaluation results of the proposed method in kitti dataset}
    \end{center}
\end{table*}

\begin{figure}
   \centering
   \includegraphics[width=1\linewidth]{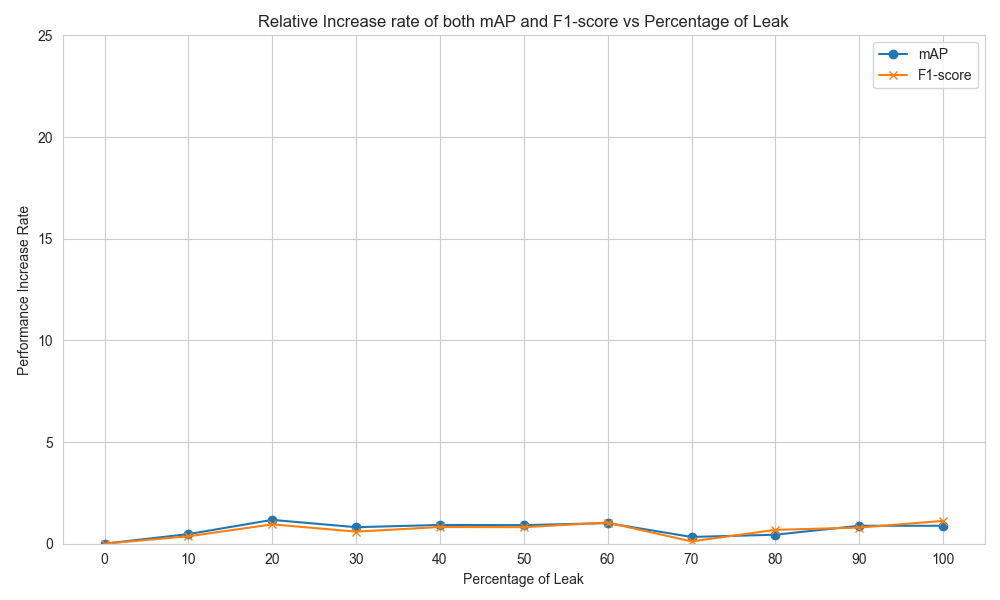}
   \caption{Relative performance increase rate for Kitti}
   \label{Fig: Relative performance increase rate for Kitti}
\end{figure}

\begin{figure}
	\centering
	\begin{subfigure}[b]{0.45\textwidth}
		\centering
		\includegraphics[width=\linewidth]{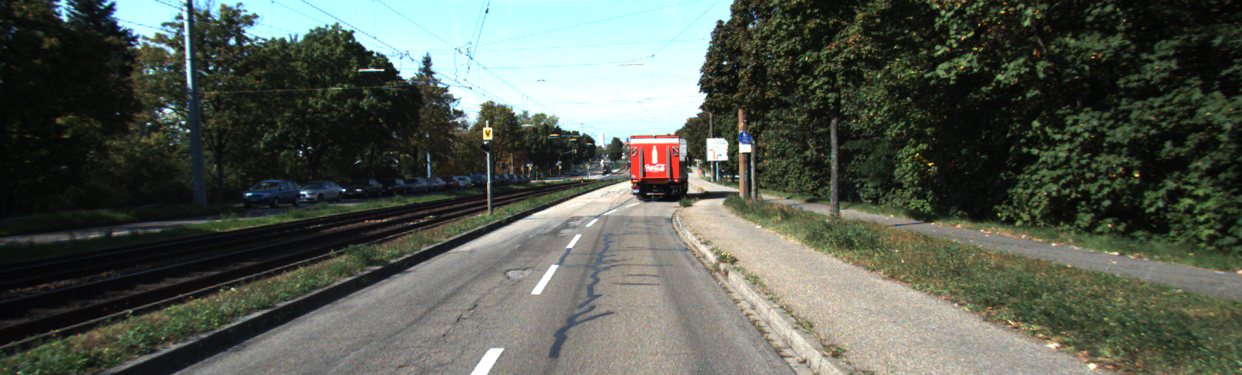}
		\caption{Train image (001453.png).}
		\label{image:image_1}
	\end{subfigure} \\
 \begin{subfigure}[b]{0.45\textwidth}
		\centering
		\includegraphics[width=\linewidth]{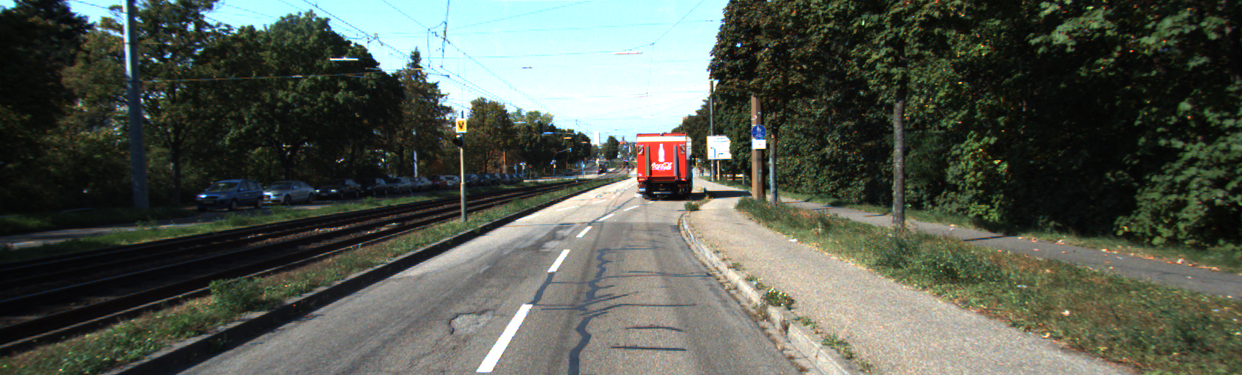}
		\caption{Test image (005442.png).}
		\label{image:image_2}
	\end{subfigure}

 \caption{Example 1 of very similar images (with pHash distance = 0) present in both train and test datasets}\label{fig:example_1}
\end{figure}

\begin{figure}
	\centering
	\begin{subfigure}[b]{0.45\textwidth}
		\centering
		\includegraphics[width=\linewidth]{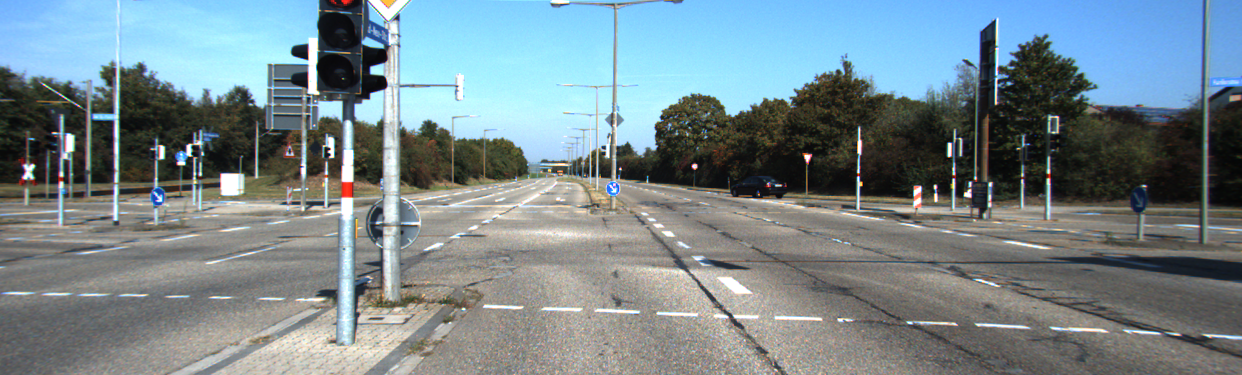}
		\caption{Train image (000017.png).}
		\label{image:image_3}
	\end{subfigure} \\
 \begin{subfigure}[b]{0.45\textwidth}
		\centering
		\includegraphics[width=\linewidth]{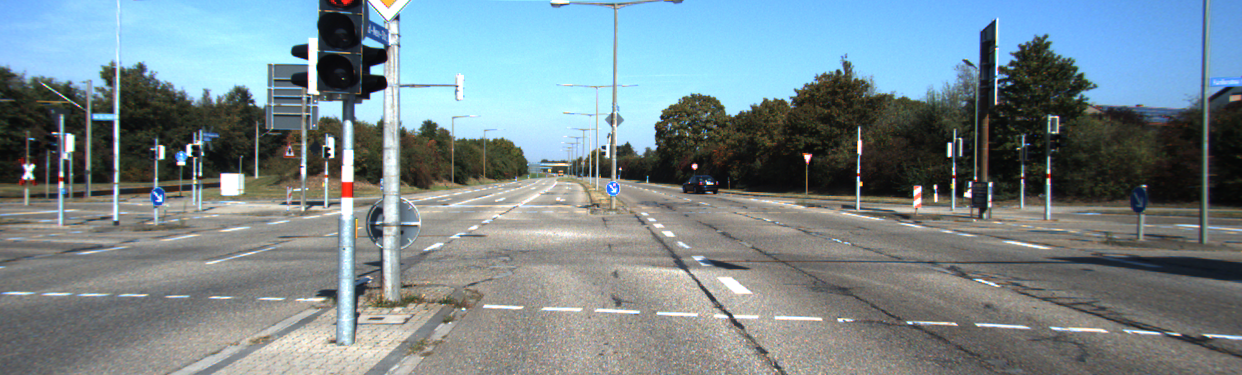}
		\caption{Test image (006279.png).}
		\label{image:image_4}
	\end{subfigure}

 \caption{Example 2 of very similar images (with pHash distance = 0) present in both train and test datasets}\label{fig:example_2}
\end{figure}

\begin{figure}
	\centering
	\begin{subfigure}[b]{0.45\textwidth}
		\centering
		\includegraphics[width=\linewidth]{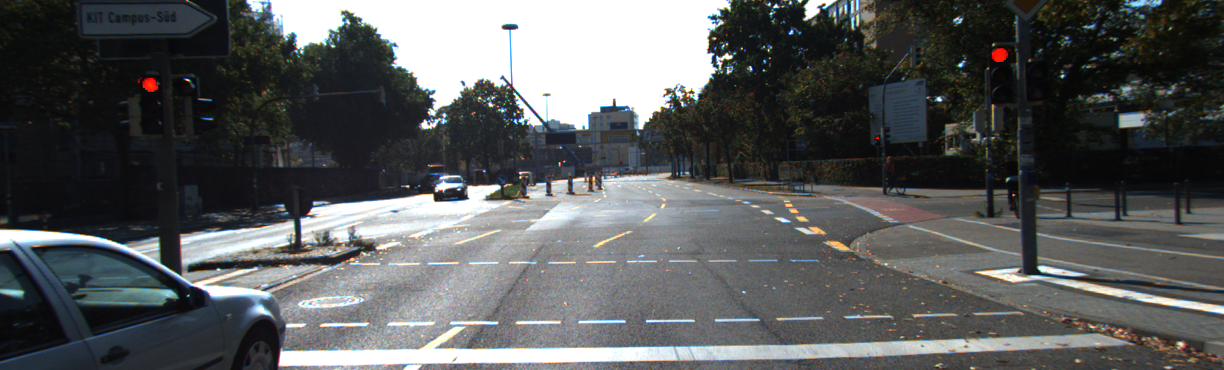}
		\caption{Train image (000771.png).}
		\label{image:image_5}
	\end{subfigure} \\
 \begin{subfigure}[b]{0.45\textwidth}
		\centering
		\includegraphics[width=\linewidth]{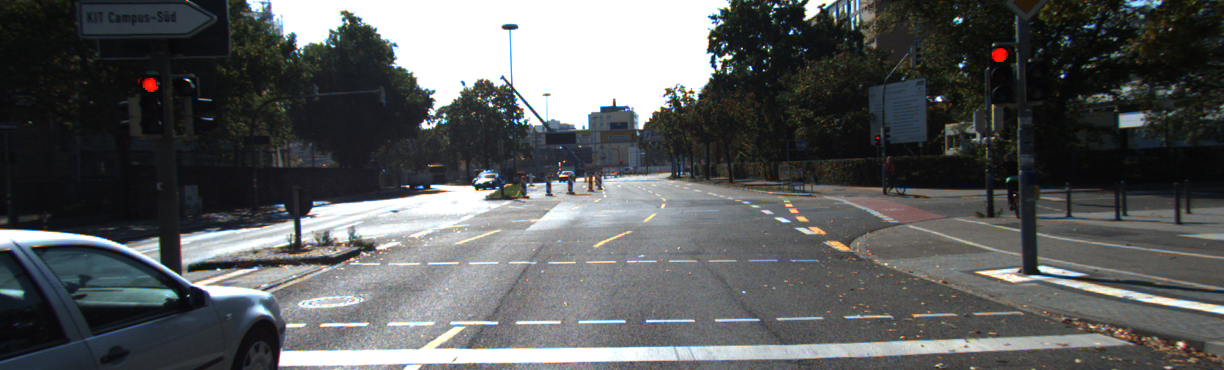}
		\caption{Test image (005790.png).}
		\label{image:image_6}
	\end{subfigure}

 \caption{Example 3 of very similar images (with pHash distance = 0) present in both train and test datasets}\label{fig:example_3}
\end{figure}

The evaluation results presented in the table show that the relative increase rate for both mAP and F1-score are lower than 5\% in every step. The increase rate is always in the range of 0-2\%. Figure \ref{Fig: Evaluation results summary on kitti} shows the pattern of increase in mAP and F1-score were not sharper for 0-20\% leakage as like the graph presented in Figure \ref{Fig: Results summary graph}. The relative performance increase rate graph shown in Figure \ref{Fig: Relative performance increase rate for Kitti} also verifies the fact that the relative increase rate never goes up by more than 1.17\% for mAP and 1.12\% for F1-score. According to our proposed method, there is a potential data leakage present in the initial split prepared from the Kitti dataset. Our manual investigation by looking into all the images verifies that there are a substantial number of cases where an image in the “train set” looks almost identical to multiple images in the “test set”. 
A few example images are presented in Figures \ref{fig:example_1}, \ref{fig:example_2}, and \ref{fig:example_3}. The images belonging to either “train” or “test” are mentioned in the sub-captions along with the actual image names/titles from the original Kitti dataset.
There are lots of similar images that can be easily spotted. For example, images \ref{image:image_1}, \ref{image:image_3}, and \ref{image:image_5} belong to the “train” dataset which are very similar or in other words almost identical to images \ref{image:image_2}, \ref{image:image_4}, and \ref{image:image_6} respectively. Even the pHash distance of those image pairs is found 0 (zero), as mentioned in the captions. Due to the presence of almost identical images in both “train” and “test” data as highlighted in Figures \ref{fig:example_1}, \ref{fig:example_2}, and \ref{fig:example_3}, the model performance gets highly inflated even from the first step (Step 0 in the results).

\begin{table}[htbp]
%\scriptsize
%\footnotesize
    \caption{pHash distances of train-test image pairs of the ‘Kitti’ dataset}
    \begin{center}
        \begin{tabular}{|c|c|}
        \hline
        \textbf{pHash distances} & \textbf{\# of Occurrences} \\
        \hline
        % Train-test ratios & 0.9880 > 0.01 \\
        % \hline
        0 & 851 \\
        %\hline
        2 & 3662 \\
        %\hline
        4 & 5141 \\
        %\hline
        6 & 6047 \\
        %\hline
        8 & 5417 \\
        %\hline
        10 & 6251 \\
        \hline
        Total & 27000 \\
        %\hline
        \hline
        \end{tabular}
        \label{tab: pHash distance of the ‘Kitti’ dataset}
    \end{center}
\end{table}

To further validate the findings, we have calculated pairwise perceptual hash (pHash) distances of all the image pairs between train and test datasets. The results presented in Table \ref{tab: pHash distance of the ‘Kitti’ dataset} also support that the ‘Kitti’ train and test images have very high similarity compared to the images of the ‘Cirrus’ dataset pHash distances as reported previously in Table \ref{tab: pHash distance of the ‘Cirrus’ dataset}. The lowest pairwise pHash distance of the ‘Kitti’ dataset is found 0 for 851 image pairs whereas the ‘Cirrus’ train-test image pairs have the lowest value of 4 and it occurs only for 2 image pairs. In addition, our findings stated in Table \ref{tab: pHash distance of the ‘Kitti’ dataset} also demonstrate that only 47 pairs of ‘Cirrus’ dataset have pHash distance of $\leq$ 10 where ‘Kitti’ has more than 27,000 such image pairs (more than 57 times higher). This clearly shows how similar the train and test images of the ‘Kitti’ dataset are.

Therefore, our conclusion is that there is a data leakage in the initial split of the Kitti dataset, which we successfully detected by applying our proposed method.

\section{Discussion}
\label{sec: discussion}
In this section, we review the findings of this study and consider their wider implications for both practitioners and researchers, aiming to detect the presence of data leakage in their existing data split(s).

%\todo[inline, color=green]{There are two things that I think are important. 1) the fact that we consider similar images to be leaked, not only identical, and 2) that this is for object detection (at least now). You can also mention that this is a novel way of detecting data leakage since no metrics exist yet.}

A method for data leakage detection has been presented based on the empirical evidence found by using one automotive industry dataset, Cirrus. Each of the sequences of Cirrus contains images from a particular road environment/scenario which can also be represented as different geographic locations. This property helps to ensure no data leakage can occur if the individual sequences are not spread over both “train” and “test” datasets. The evaluation results presented in Section \ref{sec: results} show an increasing pattern (particularly mAP and F1-score) of the model performance graph with incremental leakage of data. %This answers our RQ1 of how incremental data leakage impacts the model performance and RQ2 about which performance measure(s) helps to detect the presence of data leakage. 
However, the nature of this increasing graph is different for the first couple of steps (with the presence of 10\% and 20\% data leakage) compared to the rest of the steps with 30-100\% data leakage. This proves that introducing 10-20\% data leakage on a particular leakage-free data split can highly accelerate the model performance. On the contrary, this acceleration will not be such high if the initial split already suffers from data leakage which can be confirmed by looking at steps 3-10 in Section \ref{sec: results}.

\begin{mdframed}[style=graybox]
%\footnotesize
\scriptsize
\textbf{Answer to RQ1:} The incremental data leakage increases overall model performance in terms of all performance measures (precision, recall, mAP, and F1-score). Among them, only mAP and F1-score values consistently follow the increase pattern whereas precision and recall values often fluctuate and do not show such consistent patterns. \\
%\textbf{Answer to RQ2:} The mAP and F1-score either independently or jointly help to indicate the presence of potential data leakage in an existing data split.
\end{mdframed}

To measure how the performance of the model changes with the increased data leakage, we calculated the relative performance increase rate in each step of intentional data leakage. The numbers reported in a table in Section \ref{sec: proposed method} display the difference. Introducing data leakage to a leakage-free data split in steps 1 and 2 influenced mAP and F1-score to rise sharply by more than 12\% in every step. Nevertheless, the performance scores never rose by more than 5\% after introducing more data leakage on an existing split that is already suffering from leakage. Based on those findings, a method for data leakage detection in an arbitrary data split has been proposed in Section \ref{sec: proposed method}, and an algorithmic representation of this method is also depicted in Algorithm \ref{algo1} for better visualization of our proposed method.

\begin{mdframed}[style=graybox]
%\footnotesize
\scriptsize
\textbf{Answer to RQ2:} The presence of any potential data leakage can be detected by introducing a certain percentage of intentional leakage to an arbitrary data split and comparing the model performance by calculating the relative performance increase in each step. If the relative increase rate is found $\le$ 5\% at least once in two consecutive steps with 10\% and 20\% data leakage, then there is a high chance of having data leakage in that examined split.
\end{mdframed}

In addition, the method was also evaluated using another popular and widely used AV dataset called ‘Kitti’. The evaluation results presented in Section \ref{sec: evaluation} suggest that the data split suffers from a data leakage problem although no duplicity leakage was ensured during the initial split. The relative performance increase rate for both mAP and F1-score never goes above 2\% from step 1. For further verification, both the “train” and “test” image data of Kitti were opened and manually checked to visualize what is there in the images and how the images look like. This manual analysis indicates the fact that there are lots of images that look very similar to each other and some look almost identical by looking at the surroundings. A few examples of such almost identical images are shown in Figures \ref{fig:example_1}, \ref{fig:example_2}, and \ref{fig:example_3}. When the trained model sees such highly similar images to what it has already seen during training, the model can successfully detect all the objects which enhances the overall performance of the model. However, if the model is given to detect the same objects in a different environment than its known ones, it may struggle to do so because of the lack of proper learning about the object properties. 
This phenomenon usually makes the model memorize (instead of generalize) the correlation of the object patterns with the images' background environment instead of having a generalized learning of the object properties. This might be the reason for such high mAP (0.852) and F1-score (0.839) of the model for the Kitti dataset even from the first step (step 0) where no incremental leakage was introduced.

\begin{mdframed}[style=graybox]
\footnotesize
%\scriptsize
\textbf{Answer to RQ3:} The presence of data leakage in the initial split was successfully detected by the proposed method for data leakage detection.
\end{mdframed}

The last finding is quite significant because the Kitti dataset has been used in over 10,000 studies for training and testing of image recognition systems. The identified data leakage means that there is a systematic error in studies using Kitti for benchmarking models and that this dataset should not be used for production systems without preprocessing. This finding also shows why industrial practitioners should be aware of and always need to be extra careful about their current data splittings and model development strategies to avoid the risks of potential data leakage. Moreover, developers of our industrial partner are already being presented with these findings and they also found it useful to follow as a guideline to revise current practices of building safer, reliable, and robust object detection models for the car passengers safety and security.

\section{Threats to validity}
\label{sec: validity evaluation}
The four categories of threats to validity—conclusion, internal, construct, and external—are discussed using the paradigm developed by Wohlin et al. \cite{wohlin2012experimentation}.
\paragraph{Conclusion validity}
Concerns regarding the conclusion validity revolve around factors that can impact the capacity to arrive at an accurate judgment regarding the connections between the treatment and the results of an experiment.

Measurement reliability: The mAP and F1-score measures were utilized as the main metric in this study which may not consistently hold true. Variations in class frequencies across different experimental conditions could lead to differing average precisions (APs) for specific classes, thereby influencing the overall mAP scores. To address this concern moving forward, steps will be taken to ensure a more balanced distribution of classes across individual data splits.

Consistency in treatment implementation: Since the splits are not consistently regulated based on parameters such as class or instance counts, there may be discrepancies in class distributions among the splits, potentially impacting performance. This issue remained unaddressed in the current experiment but will be taken into account when designing future experiments.

\paragraph{Internal validity}
Threats to internal validity pertain to factors that could potentially influence the causality of the independent variable without the researcher's awareness, thereby compromising the ability to conclusively establish a cause-and-effect relationship between the treatment and the observed outcome.

Maturation: One such threat is maturation, which arises when the OD model is trained for 100 epochs for all the steps, regardless of any considerations regarding loss or accuracy thresholds. This practice may introduce variability in data points among the steps, thereby posing a risk to internal validity. To address this concern, the model was trained for 500 epochs in the majority of steps, yet the observed increase in performance measures was found to be non-significant, ranging between 0.006 and 0.009.

%Instrumentation: Another potential threat is instrumentation, which occurs due to the use of 2D labels generated with the inference of a pre-trained model in this experiment. This introduces a risk of inconsistency in measurement tools. However, this threat was mitigated by manually inspecting and correcting any inaccuracies in the drawn bounding boxes on all images.

\paragraph{Construct validity}
Construct validity refers to the degree to which the outcomes of an experiment can be generalized or applied to the fundamental concept or theory that underpins the experiment.

Mono-operation bias: One aspect to consider is mono-operation bias, where the experiment exclusively focuses on the OD task and utilizes a related dataset and performance metric to assess the presence of data leakage. This approach may introduce a bias towards a single operation. To address this concern, future experiments will include additional operations such as image classification to explore the effectiveness of the proposed method for detecting data leakage.

Mono-method bias: Another consideration is mono-method bias, which arises from the reliance on a single method of measurement. In this case, the relative increase rate is the only method for data leakage detection which might not be always considering the varying quality and complexity of the image datasets. Future experimentation on other automotive datasets including the open source public datasets will not only help to generalize the proposed method but also to avoid this threat.

Inadequate preoperational explication of constructs: A potential threat to the construct validity of this study is the inadequate preoperational explication of constructs, particularly regarding the selection of the similarity threshold for perceptual hashing (pHash). The Hamming distance of up to 10 was used to identify similar images between datasets, but this threshold may not fully capture all degrees of similarity, potentially impacting the accuracy of conclusions about data leakage. A more detailed justification or sensitivity analysis could help align the operational definition of “similarity” with the research objectives.
%Another consideration is mono-method bias, which arises from the reliance on a single method of measurement. In this case, the calculation of mAP using the confusion matrix, intersection over union (IoU), precision, and recall may disproportionately emphasize certain aspects, potentially biasing the results. To mitigate this, alternative measures like the f1-score, which equally weigh precision and recall, could be incorporated in future experiments. Despite the inherent trade-off between precision and recall, forthcoming studies on data splits and identifying data leakages will employ both mAP and f1-score to formulate robust hypotheses.

%Furthermore, confounding constructs and levels of constructs are evident in the experiment due to variations in data splitting methods across different treatments, leading to discrepancies in the number of images in training and validation/test sets, particularly noticeable in similarity/dissimilarity-based splits. This variability stems from the utilization of unsupervised clustering methods, which were not controlled.

\paragraph{External validity}
Factors impacting external validity encompass conditions that limit our ability to extrapolate the results of our experiment to real-world industrial contexts.

The interaction of setting and treatment poses a potential external threat, as the experiment solely utilizes the YOLOv7 object detection model. Different 2D object detection models may yield disparate performance scores. However, the experiment's scope did not encompass the exploration of alternative models. The literature referenced in the study indicates that the YOLOv7 model was chosen for its superior performance and speed, thus justifying its selection.

Similarly, the interaction of selection and treatment raises concerns regarding the class imbalance within the Cirrus dataset as well as in the Kitti dataset, which could impact the validity of the findings. While achieving perfect balance in datasets for image recognition and specifically for OD tasks is challenging, many popular benchmark datasets exhibit imbalances. To enhance the generalizability of the study's findings, future experiments could replicate the study using datasets with comparatively less imbalance.

%Furthermore, the interaction of history and treatment introduces an external validity threat stemming from the use of a single dataset obtained from an automotive OEM. Expanding the dataset pool is a focal point of ongoing research efforts, aimed at mitigating this limitation and enhancing the study's external validity.
\section{Conclusion and future work}
\label{sec: conclusion}

%\todo[inline, color=green]{In the conclusions, it is very important that you think like a software engineer at Volvo -- that this increases the safety of modern cars, because it provides more truthful measure of network performance. This means that we don't stop training pre-maturely and therefore reduce the risks of releasing cars that are not safe. }
Data leakage detection is very important particularly in the context of automotive perception systems to ensure the safety of the passengers. Detecting data leakage in ML models, particularly in tasks like OD, is crucial not just for improving model accuracy but also for ensuring the integrity and reliability of software systems as a whole. Failure to detect data leakage during training and deployment of an OD model may put risks of deploying an incorrectly and insufficiently trained model which would fail to correctly detect and classify objects in unseen scenarios. This is why automotive OEMs, like Volvo Cars, put high emphasis on detecting and removing any form of data leakage prior to training the model in order to ensure safe and secure models to be deployed on cars.

In the broader field of software engineering (SE), the verification and validation (V\&V) of ML components are directly tied to software quality. Data leakage, if undetected, can lead to overfitting during model development, resulting in ML models that perform well in training but fail under real-world conditions. This directly impacts the reliability of software systems that integrate these models, leading to poor decision-making, degraded user experience, and compromised system performance in real-time deployment. %Addressing data leakage early on enhances the robustness of the entire software lifecycle, ensuring that the final product meets both functional and non-functional requirements in SE, such as scalability, maintainability, and dependability.

In this study, a method for data leakage detection is proposed. The proposed method is based on the empirical results of experiments conducted with the “Cirrus” dataset. According to the proposed method, if the model performance does not increase by more than 5\% after successively leaking at least 20\% of the “test” data to “train” data (10\% in every step), then there is a high chance of potential data leakage in the existing data split. This method was evaluated on one of the most popular benchmark datasets called “Kitti” to verify whether data leakage can be detected or not by this method. The evaluation results indicated the presence of potential data leakage presence in the initial split of the Kitti dataset. This finding was further justified by individually looking into the images in the “train” and “test” datasets and finding highly similar-looking images as shown in some examples in Figures \ref{fig:example_1}, \ref{fig:example_2}, and \ref{fig:example_3}.

However, this method needs to be re-evaluated on other automotive datasets. So we are planning to test the method in the future not only on the publicly available datasets but also on real in-use image datasets used by our industrial partner for training such models. Some other OD models including the industrial in-use are also planned to be used to validate the method as the next step. Future work also includes finding metrics to measure data leakage upon successful detection.

% %%
% %% The next two lines define the bibliography style to be used, and
% %% the bibliography file.
\bibliographystyle{IEEEtran}
\bibliography{author-draft}
\end{document}